\documentclass[10pt,tightenlines,eqsecnum,floats,aps,nofootinbib,prd]{revtex4-2}
\usepackage{amsmath,amssymb,amsfonts,amsthm,amscd}
\usepackage{graphicx}
\usepackage{amsmath}
\usepackage{amssymb}
\usepackage{graphicx}
\usepackage{cancel}
\usepackage{slashed}
\usepackage{mathtools}
\usepackage{tcolorbox}
\usepackage{tikz-cd}
\usepackage{tikz}
\usepackage{mathrsfs}
\usepackage{relsize}
\usepackage{natbib}
\usepackage[margin=1in]{geometry}
\usepackage{hyperref}
\usepackage[normalem]{ulem}

\newtheorem{lemma}{Lemma}
\usetikzlibrary{shapes.geometric, arrows}
\tikzstyle{startstop} = [rectangle, rounded corners, minimum width=3cm, minimum height=1cm,text centered, draw=black, fill=red!30]
\tikzstyle{arrow} = [thick,->,>=stealth]
\begin{document}
\title{Towards a Connection Formulation of Action-Dependent Palatini Gravity}
\author{Callum Bell}
\email{c.bell8@lancaster.ac.uk}
\affiliation{Department of Physics, Lancaster University, Lancaster UK}

\author{David Sloan}
\email{d.sloan@lancaster.ac.uk}
\affiliation{Department of Physics, Lancaster University, Lancaster UK}

\begin{abstract}
    Previous work has shown that first-order Palatini gravity may be cast as a non-conservative Herglotz Lagrangian field theory, which excludes the conformal mode from its space of dynamical variables. In the present work, we analyse the role of the $SO(1,3)$ gauge connection in the construction of this action-dependent theory. The connection may be algebraically decomposed into complementary sectors which we interpret as `shape' and `scale' components. Only the former is retained within the Herglotz theory, while the latter is incorporated into the action-dependent sector. It is shown that the choice of connection decomposition is a space that can be smoothly parameterised. Such an observation leads to the finding that there exists a `moduli space of frictional field theories', in which each point corresponds to a distinct algebraic splitting of the connection into shape and scale pieces. Movement within the moduli space transitions between Herglotz Lagrangians whose on-shell dynamics all reproduce first-order Palatini gravity. The distinction lies in how the reproduction of the scale dynamics of the original theory is partitioned between action dependence and dynamical torsion.
\end{abstract}
\maketitle
\section{Introduction}\label{Sec:Introduction}
In previous work \cite{bell2026classical}, it has been demonstrated that the gauge-invariant observable dynamics of the first-order Palatini Lagrangian are impervious to rescalings of the conformal mode $\phi$. This was achieved decomposing the coframe $e^I$ into the product $e^I=e^{\phi/2}\tilde{e}^I$, in which $\tilde{e}^I$ denotes a coframe of fixed unit determinant. The action was then cast in a form which made manifest the fact that the vector field $\partial_\phi$ possesses all the requisite properties of a dynamical similarity \cite{sloan2018dynamical,sloan2021scale,bravetti2023scaling}. Following the procedure developed in \cite{DSMulti} for the symmetry reduction of multisymplectic Lagrangian field theories, we obtained an action-dependent Herglotz Lagrangian for the Palatini theory, which was of an inherently non-conservative nature \cite{cannarsa2019herglotz}. The physical interpretation of this construction was as follows. \\

The conformal mode is not an empirically accessible degree of freedom, and further, is not fundamental to the dynamical closure of the algebra of gauge-invariant observables. It therefore represents superfluous structure within the spacetime metric \cite{gryb2021scale,ismael2003symmetry,ismael2021symmetry,van2007representation}. While not, a priori, an immediate problem, we have several well-motivated arguments which disfavour the presence of superfluous structure. Perhaps the most physically compelling of these is the possibility that the superfluous structure degenerates, while the underlying physical description remains perfectly well-defined. Examples of such a phenomenon include simple cosmological models (such as FLRW and Bianchi class-A universes) where the degeneration of a scale factor renders the theory non-predictive at the initial singularity. Recent work suggests that these points, typically considered to mark the classical breakdown of general relativity, admit a smooth continuation, once the degenerate superfluous structure has been suitably excised \cite{sloan2019scalar,koslowski2018through}. Similarly, the definition of a Euclidean path integral for standard Einstein-Hilbert gravity is made somewhat intractable by the fact that, amongst other things, the integrand is unbounded from below \cite{gibbons1978path}. In particular, arbitrarily large variations of the conformal mode over the space of metrics drive the Euclidean action to be arbitrarily negative.\\

By eliminating the conformal mode, these (and other) problems are, in the most favourable of cases, resolved, or at the very least lessened in severity. $\phi$ encodes a local choice of scale at each spacetime point. The Herglotz Lagrangian, which makes no reference to $\phi$, therefore has no manifest notion of scale. While the conformal mode does not contribute to the dynamical evolution of the observables, this does not imply there exists no `scale-related physics'. Indeed, it is precisely for this reason that the symmetry-reduced theory is action-dependent! Were the conformal mode to be irrelevant for physics as a whole, then upon excising $\phi$ from our description, only the theory of unimodular gravity would remain, which is \textit{not} the case.\\

The physical interpretation is that within the dynamics of the original Palatini theory, we may identify two distinct sectors. These roughly correspond to the evolution of the conformal geometry, and dynamical changes in local scale. We separate these schematically into dynamics of `shape $+$ scale' degrees of freedom. Within the Herglotz theory, the scale sector is no longer present. Instead, the observable dynamics are reproduced by a theory in which the shape degrees of freedom experience friction-like interactions, encoded within the action dependence of the Herglotz Lagrangian \cite{de2023multicontact,de2025practical}. The qualifier `friction-like' is used to highlight that the non-conservative behaviour need not be inherently energy-reducing. The intuitive picture is therefore that the empirical findings of a spacetime observer are completely insensitive to whether we consider the full Palatini theory of gravity, complete with its local choice of scale encoded by the conformal mode, or a modified unimodular theory, in which local scale physics has been replaced with non-conservative behaviour.\\

These were the physical conclusions extracted from the work in \cite{bell2026classical}. The aim of the present article is to demonstrate that this construction, and subsequent physical interpretation, is but a particular case of a significantly richer mathematical story. We demonstrate that the task of reproducing the scale dynamics is typically not assigned exclusively to the action-dependent sector. Indeed, in the most general of cases, we find that a combination of action dependence and torsion-like properties provides the mechanism capable of compensating the elimination of the conformal mode. In section (\ref{Sec:Decomp}), we begin with a systematic approach to the decomposition of the space of connection values. This is an affine space modeled on a real 24-dimensional vector space. This space is split into a direct sum of shape and scale components using a surjective linear map.\\

Sections (\ref{Sec:Expansion}) and (\ref{Sec:Herglotz}) then use this decomposition to compute the Herglotz Lagrangian, in accordance with the techniques of \cite{DSMulti}. The dynamical content of this Herglotz Lagrangian is the subject of section (\ref{Sec:EoM}); it is found that, on-shell, the theory exhibits torsion-like effects. An extended discussion of the role of this behaviour is presented in section (\ref{Sec:Torsion}).\\

There are two important and related observations to be made following the results of sections (\ref{Sec:Decomp})-(\ref{Sec:Torsion}). The first is that the functional form of the Herglotz Lagrangian (\ref{Eq:HerglotzL2}) obtained in section (\ref{Sec:Herglotz}) differs to that of the Lagrangian deduced in \cite{bell2026classical}. The second is that the difference between the two approaches is limited to the way in which the gauge connection is decomposed. These observations strongly suggest that the manifestation of the torsional and frictional characteristics are decomposition-dependent. Further, we are motivated to conjecture the existence of a means of parameterising the possible choices of connection decomposition. This proposal is made mathematically concrete in section (\ref{Sec:Abstraction}), and in section (\ref{Sec:Nemo}), we provide a group-theoretic argument that not only validates the proposal, but also shows that the resulting parameterisation is continuous.\\

It is found that the Herglotz theory describing first-order Palatini gravity is in fact characterised by a one-dimensional moduli space. Each point of the moduli space is an equivalence class of surjective bundle morphisms, which provide a pointwise decomposition of the space of local connection values. Physically, different points in the moduli space correspond to distinct ways to project the spin connection onto disjoint shape and scale sectors. Each decomposition leads to a Herglotz Lagrangian in which the burden of reproducing the scale dynamics is partitioned in a characteristic manner between torsional and action-dependent sectors. The moduli space construction is explored extensively in sections (\ref{Sec:ModuliSpace})-(\ref{Sec:Bundle}). Finally, we conclude with a summary of our results, together with a number of proposed directions for further work.
\section{Decomposition of the Connection Space}\label{Sec:Decomp}
We work on a four-dimensional, oriented, smooth manifold $M$, which we take to be closed and compact. Let $g$ be a Lorentzian metric on $M$, and denote by $SO(M)$ its bundle of oriented orthonormal frames, with structure group $SO(1,3)$. Let $E:=SO(M)\times_{SO(1,3)}V$ be the associated internal Minkowski vector bundle, where $V\cong\mathbb{R}^{1,3}$ is equipped with the Minkowski metric $\eta=\text{diag}(-1,1,1,1)$. A non-degenerate coframe at $x\in M$ is a linear isomorphism $e_x:T_xM\longrightarrow E_x$, or as we shall write for simplicity, an element $e_x^I\in T^*_xM\otimes V$. To isolate the conformal scaling mode of gravity, the coframe field is decomposed as $e^I=e^{\phi/2}\tilde{e}^I$, with $\text{det}(\tilde{e}_\mu^{\;I})=1$. The kinematic background is governed by the unimodular area $2$-form
$\tilde{\Sigma}_{IJ}\in\Gamma(\Lambda^2T^*M\otimes\Lambda^2V^*)$, defined via (see appendix \ref{App:A} for differential form conventions)
\begin{equation}\label{Eq:Sigma}
    \tilde{\Sigma}_{IJ} =\frac{1}{2}\varepsilon_{IJKL}\,\tilde{e}^K\wedge\,\tilde{e}^L = \star\;(\tilde{e}_I\wedge\tilde{e}_J)
\end{equation}
At a fixed point $x\in M$, the fibre $\mathcal{A}_x$ of the affine bundle of connections is an affine space, modeled on the finite-dimensional vector space
\begin{equation}\label{Eq:ConnSp}
    U_x:=T_x^*M\otimes\Lambda^2V
\end{equation}
which is of dimension $4\times6=24$. The space of localised $3$-forms is $W_x:=\Lambda^3T_x^*M$, which has dimension $4$. The objective is to introduce, at each point $x\in M$, a surjective linear map $\Phi^{(1)}$, which takes an element $\omega_x$ of the finite-dimensional vector space $U_x$, and returns a local spacetime 3-form $S\in W_x$. The precise representation-theoretic justification of this construction will be given below.\\

The particular map $\Phi^{(1)}$ we introduce serves to extract the trace-like scale momentum from the 24 components of $\omega_x$. It is defined as
\begin{equation}\label{Eq:Phi1}
    \begin{split}
        \Phi^{(1)} \;:\; U_x \;&\longrightarrow\; W_x\\
        \omega_x\;&\longmapsto \; \Phi^{(1)}(\omega_x):=-\tilde{\Sigma}_{IJ}\wedge \omega_x^{IJ}
    \end{split}
\end{equation}
From the dimensions of the spaces involved, it is clear that $\text{dim\,Ker}(\Phi^{(1)})=20$.\\

Since $\Phi^{(1)}$ is surjective, there exists a right inverse, $\Psi^{(1)}$, which lifts an element $S\in W_x$ into the model space $U_x$. It is straightforward to verify that, with $\sigma:=\star\, S$, the correct expression reads
\begin{equation}\label{Eq:Psi1}
    \begin{split}
        \Psi^{(1)} \;:\;  W_x \;&\longrightarrow\;T^*_xM\otimes \Lambda^2V\\
        S\;&\longmapsto \; \Psi^{(1)}(S):= \frac{1}{3}\tilde{e}^{[I}\iota^{J]}\sigma
    \end{split}
\end{equation}
where $\iota^I:=\iota_{\tilde{e}^I}$ denotes the interior product with respect to the coframe field. Similar notation will be employed for contraction with the frame vector dual to $\tilde{e}^I$: $\iota_I:=\iota_{\tilde{e}_I}$.\\

A direct calculation shows that $\Phi^{(1)}\circ\Psi^{(1)}=\text{id}$, and that there is no intersection between $\text{Ker}(\Phi^{(1)})$ and $\text{Im}(\Psi^{(1)})$. From the dimensionality of these spaces, the rank-nullity theorem guarantees that vector space $U_x$ decomposes uniquely as the following direct sum
\begin{equation}\label{Eq:DirectSum}
    U_x=\text{Ker}(\Phi^{(1)})\oplus\text{Im}(\Psi^{(1)})
\end{equation}
The existence of a canonical splitting of the model vector space of connection differences at $x$ allows us to uniquely decompose $\omega_x$ into a `shape' piece, encoded in $\text{Ker}(\Phi^{(1)})$, and a `scale' piece, determined by $\text{Im}(\Psi^{(1)})$. Thus, we write $\omega_x=(\omega_\perp)_x+(\omega_s)_x$, in which $(\omega_\perp)_x$ refers to the shape component. Note that this is strictly a decomposition of the model space $U_x$, and not of the space of connections. The latter is an infinite-dimensional affine space modeled on the space of $\frak{so}(1,3)$ valued 1-forms $\Omega^1(M,\frak{so}(1,3))$. The former, by contrast is a genuine, finite-dimensional vector space $T^*_xM\otimes \Lambda^2V$, which may perfectly well be decomposed as the direct sum (\ref{Eq:DirectSum}).\\

A local $SO(1,3)$ gauge transformation with parameter $\Lambda$ acts on $\mathcal{A}_x$ inhomogeneously:
\begin{equation}
    \omega_x\,\longrightarrow\, \Lambda^{-1}(x)\omega_x\Lambda(x) + \Lambda^{-1}(x)d\Lambda(x)
\end{equation}
Since $\Psi^{(1)}$ is the right inverse of $\Phi^{(1)}$, we may introduce the operator $P:=\Psi^{(1)}\circ \Phi^{(1)}$, which acts on $U_x$ and projects onto the scale component. Having chosen a local trivialisation, and hence an affine identification of the fibre with its model vector space after selecting an origin, we may apply the linear splitting of $U_x$ to the corresponding connection (difference). The resulting quantities $\omega_\perp$ and $\omega_s$ are algebraically defined projected components, but neither will, in general, transform as a connection 1-form. Instead, we have 
\begin{equation}\label{Eq:Tranf}
    \omega_\perp\,\longrightarrow\, \Lambda^{-1}\omega_\perp \Lambda + (1-P)(\Lambda^{-1}d\Lambda),\qquad \omega_s\,\longrightarrow\, \Lambda^{-1}\omega_s \Lambda + P(\Lambda^{-1}d\Lambda)
\end{equation}
Consequently, it is the inhomogeneous affine translation $\Lambda^{-1}d\Lambda$ that presents the obstruction to the individual pieces $\omega_\perp$ and $\omega_s$ defining connections. We introduce the object $D_\perp:=d+\omega_\perp$, in which $d$ denotes the usual exterior derivative. Note that $D_\perp$ is not a gauge-covariant derivative; it is an algebraically useful derivation induced by the projected component $\omega_\perp$, and satisfies the graded Leibniz rule.
\section{Expansion of the Palatini Action}\label{Sec:Expansion}
Having introduced the pointwise splitting of the model vector spaces $U_x$ using the maps $\Phi^{(1)}$ and $\Psi^{(1)}$, we now substitute the algebraic decomposition $\omega=\omega_\perp+\omega_s$ directly into the $2$-form field strength, given by $R^{IJ}(\omega)=d\omega^{IJ}+\omega^I{}_K\wedge\omega^{KJ}$. This gives
\begin{equation}
    R^{IJ}(\omega) = \underbrace{ \left(d\omega_\perp^{IJ} +\omega_\perp^I{}_K\wedge\omega_\perp^{KJ}\right)}_{:=\,R_\perp^{IJ}}\;\, + \;\, \underbrace{\left(d\omega_s^{IJ}+\omega_\perp^I{}_K\wedge\omega_s^{KJ} +\omega_s^I{}_K\wedge\omega_\perp^{KJ}\right)}_{=\,D_\perp\omega_s^{IJ}}\;\,+\;\, \omega_s^I{}_K\wedge\omega_s^{KJ}
\end{equation}
where we have grouped the terms according to their dependence on $\omega_s$, and have identified the derivation $D_\perp$ defined above. Due to the transformation properties of $\omega_\perp$, $R^{IJ}_\perp$ is an algebraic component of $R^{IJ}$, and is not itself the curvature of a connection. The notation employed is symbolically reminiscent of the usual tensorial objects, but this is the extent of the analogy. We shall see in due course the consequences of this non-covariant behaviour.\\

With this, the curvature $R^{IJ}$, which is a genuine $SO(1,3)$ tensor, cleanly factors as follows
\begin{equation}\label{Eq:FactR}
    R^{IJ}(\omega)=R_\perp^{IJ}+D_\perp\omega_s^{IJ}+\omega_s^I{}_K\wedge\omega_s^{KJ}
\end{equation}
in which we emphasise that this is only an algebraic identity. The decomposed curvature tensor is now substituted into the first-order Palatini Lagrangian \cite{peldan1994actions} (we do not bother to keep track of overall numerical prefactors) 
\begin{equation}
    \mathcal{L}=\Sigma_{IJ}\wedge R^{IJ}(\omega) =e^\phi\,\tilde{\Sigma}_{IJ}\wedge\left(R_\perp^{IJ}+D_\perp\omega_s^{IJ}+\omega_s^I{}_K\wedge\omega_s^{KJ}\right)
\end{equation}
Making use of the graded Leibniz rule, the term linear in $\omega_s$ may be rearranged as follows
\begin{equation}
    \tilde\Sigma_{IJ}\wedge D_\perp\omega_s^{IJ}=D_\perp\left(\tilde\Sigma_{IJ}\wedge\omega_s^{IJ}\right)-\left(D_\perp\tilde\Sigma_{IJ}\right)\wedge\omega_s^{IJ}
\end{equation}
Since $\tilde\Sigma_{IJ}\wedge\omega_s^{IJ}$ has all internal $SO(1,3)$ indices contracted, we may make the replacement $D_\perp\rightarrow d$. This is an algebraic property that holds irrespective of the fact that $\tilde\Sigma_{IJ}\wedge\omega_s^{IJ}$ does not transform as a scalar, and that $D_\perp$ is not a gauge-covariant derivative. Hence, up to boundary terms, we have
\begin{equation}\label{Eq:FinalL}
    \mathcal{L} =e^\phi\,\Big[\tilde{\Sigma}_{IJ}\wedge R_\perp^{IJ}-d\phi\wedge\left(\tilde{\Sigma}_{IJ}\wedge\omega_s^{IJ}\right)-\left(D_\perp\tilde{\Sigma}_{IJ}\right)\wedge\omega_s^{IJ}+\tilde{\Sigma}_{IJ}\wedge\omega_s^I{}_K\wedge\omega_s^{KJ}\Big]
\end{equation}
\section{The Herglotz Lagrangian}\label{Sec:Herglotz}
The Lagrange density (\ref{Eq:FinalL}) is now of the form to which the techniques of \cite{DSMulti} may be applied. As developed in appendix \ref{App:B}, the reduction process may be generalised slightly for those cases in which the Lagrange density is expressed as $\mathcal{L}=e^\phi\mathcal{F}$, for some top-form $\mathcal{F}$. This is precisely the case of (\ref{Eq:FinalL}), and using the definition (\ref{Eq:DefS}) of the action 3-form, we have
\begin{equation}\label{Eq:HerglotzS}
    S = \frac{\partial \mathcal{F}}{\partial(d\phi)}= -\left(\tilde\Sigma_{IJ}\wedge\omega_s^{IJ}\right)
\end{equation}
and hence the Herglotz Lagrangian density $\mathcal{L}^H=\mathcal{F}-d\phi\wedge S$ is
\begin{equation}\label{Eq:HerglotzL1}
    \mathcal{L}^H=\tilde{\Sigma}_{IJ}\wedge R_\perp^{IJ}-\left(D_\perp\tilde{\Sigma}_{IJ}\right)\wedge\omega_s^{IJ}+\tilde{\Sigma}_{IJ}\wedge\omega_s^I{}_K\wedge\omega_s^{KJ}
\end{equation}
The final step is to replace all explicit instances of $\omega_s^{IJ}$ with the $3$-form $S$, which is achieved by inverting the relation (\ref{Eq:HerglotzS}). The final term is simple, and using (\ref{Eq:Psi1}), we find
\begin{equation}
    \tilde{\Sigma}_{IJ}\wedge\omega_s^I{}_K\wedge\omega_s^{KJ}=\frac{1}{6}\sigma\wedge\star\,\sigma=\frac{1}{6} S\wedge\star\,S
\end{equation}
The second term of (\ref{Eq:HerglotzL1}) is a torsion-like piece, and may be re-expressed as
\begin{equation}
    \left(D_\perp\tilde{\Sigma}_{IJ}\right)\wedge\omega_s^{IJ}=\frac{2}{3} \iota_I D_\perp\tilde e^I\wedge S
\end{equation}
We define the torsion-like contribution of the shape sector, together with its corresponding trace as
\begin{equation}\label{Eq:Torsion}
    \Theta^I:=D_\perp\tilde{e}^I,\quad\quad\quad\Theta:=\iota_I\Theta^I
\end{equation}
With this, the final form of the Herglotz Lagrangian that we obtain from the reduction of the first-order Palatini action is given by
\begin{equation}\label{Eq:HerglotzL2}
    \boxed{\mathcal{L}^H = \tilde{\Sigma}_{IJ}\wedge R_\perp^{IJ} - \frac{2}{3}\Theta\wedge S +\frac{1}{6}S\wedge\star\, S}
\end{equation}
As we have been careful to highlight throughout, objects constructed from the shape projection $\omega_\perp$ do not generally transform as tensors or connections. As such, the above Herglotz Lagrangian cannot naively be interpreted as unimodular Palatini theory with torsion, that is coupled to an action-dependent sector. The correct interpretation is slightly more subtle, and is deduced by examining how the projected variables behave under the full internal $SO(1,3)$ symmetry group, which is the task to which we now turn.
\section{Transformation Properties of $\mathcal{L}^H$}\label{Sec:Trans}
As described above, the projected components $\omega_\perp$ and $\omega_s$ do not generally transform as connection 1-forms. Referring to the expressions (\ref{Eq:Tranf}), in order to alleviate algebraic complexity in the intermediate steps, we shall denote $(1-P)(\Lambda^{-1}d\Lambda):=\alpha_\perp$, and $P(\Lambda^{-1}d\Lambda):=\alpha_s$. With this, we have that under a local Lorentz transformation
\begin{equation}\label{Eq:Trans2}
    \omega_\perp\,\longrightarrow\, \Lambda^{-1}\omega_\perp \Lambda + \alpha_\perp,\qquad \omega_s\,\longrightarrow\, \Lambda^{-1}\omega_s \Lambda + \alpha_s    
\end{equation}
We also know that, under this same $SO(1,3)$ transformation, the coframe $\tilde{e}^I$ and unimodular area 2-form $\tilde{\Sigma}_{IJ}$ behave according to
\begin{equation}\label{Eq:TransSigma}
    \tilde{e}^{\prime I}=(\Lambda^{-1})^I_J\tilde{e}^J,\qquad \; \tilde{\Sigma}_{IJ}^\prime = \Lambda_I^K\Lambda_J^L\tilde{\Sigma}_{KL}
\end{equation}
The transformation of the algebraic quantity $R^{IJ}_\perp$ contains the usual homogeneous piece, together with the additional terms generated by $\alpha_\perp$
\begin{equation*}
    R_\perp^{\prime} = \Lambda^{-1}R_\perp \Lambda + D_\perp^\prime\alpha_\perp +\alpha_\perp\wedge\alpha_\perp
\end{equation*}
in which $D_\perp^\prime$ is the derivation formed from the transformed $\omega_\perp^\prime$. This, together with the transformation (\ref{Eq:TransSigma}) implies that the product $\tilde{\Sigma}_{IJ}\wedge R^{IJ}_\perp$ acquires non-trivial contributions from $\alpha_\perp$. Concretely, we find that
\begin{equation}\label{Eq:TransSigmaR}
    \tilde{\Sigma}^\prime_{IJ}\wedge R^{\prime IJ}_\perp = \tilde{\Sigma}_{IJ}\wedge R^{ IJ}_\perp + \tilde{\Sigma}_{IJ}\wedge D^\prime_\perp\alpha_\perp^{IJ} + \tilde{\Sigma}_{IJ}\wedge \alpha^I_{\perp K}\wedge \alpha_\perp^{KJ}
\end{equation}
Consider now the action 3-form $S$, which we recall is given by
\begin{equation}
    S=-\tilde{\Sigma}_{IJ}\wedge \omega^{IJ}_s
\end{equation}
While this expression has all internal indices contracted, it is not gauge invariant under internal $SO(1,3)$ transformations. From (\ref{Eq:Trans2}) and (\ref{Eq:TransSigma}), it is easy to see that
\begin{equation}\label{Eq:TransS}
    S^\prime=S -\tilde{\Sigma}_{IJ}^\prime\wedge\alpha_s^{IJ}
\end{equation}
The final element from which our Herglotz Lagrangian is constructed is the torsion-like piece $\Theta=\iota_I\Theta^I$. Recalling that $\Theta^I:=D_\perp\tilde{e}^I$, the transformed object $\Theta^{\prime I}$ is given by
\begin{equation}\label{Eq:TransTheta}
    \Theta^{\prime I}=(\Lambda^{-1})^I_J\Theta^J + \alpha^I_{s \,J}\wedge \tilde{e}^{\prime J}\qquad\implies\qquad \Theta'=\Theta + \iota_{I}^{\prime}\left( \alpha^I_{s \,J}\wedge \tilde{e}^{\prime J}\right)
\end{equation}
in which $\iota^\prime_I:=\iota_{\tilde{e}^{\prime}_{I}}$. Thus, we see that the torsion-like term $\Theta$ acquires an inhomogeneous contribution under local Lorentz transformations, making manifest its non-tensorial nature.\\

Recall, from section (\ref{Sec:Herglotz}), that the Herglotz Lagrangian, before eliminating the scale component $\omega_s^{IJ}$, could be expressed as
\begin{equation}\label{Eq:HerglotzL1Again}
    \mathcal{L}^H=\tilde{\Sigma}_{IJ}\wedge R_\perp^{IJ}-\left(D_\perp\tilde{\Sigma}_{IJ}\right)\wedge\omega_s^{IJ}+\tilde{\Sigma}_{IJ}\wedge\omega_s^I{}_K\wedge\omega_s^{KJ}
\end{equation}
Using the graded Leibniz rule on the second term, this may alternatively be expressed as
\begin{equation*}
    \mathcal{L}^H = \tilde{\Sigma}_{IJ}\wedge\left(R^{IJ}_\perp + D_\perp\omega_s^{IJ} +\omega^I_{s\,K}\wedge\omega_s^{KJ}\right) - d\left(\tilde{\Sigma}_{IJ}\wedge \omega_s^{IJ}\right)
\end{equation*}
in which we have used the fact that all internal indices of the term $\tilde{\Sigma}_{IJ}\wedge \omega_s^{IJ}$ are contracted in order to replace $D_\perp\rightarrow d$. The term in parenthesis that is wedged against the unimodular area 2-form $\tilde{\Sigma}_{IJ}$ is immediately recognised as the decomposition (\ref{Eq:FactR}) of the full curvature $R^{IJ}(\omega)$. We therefore write
\begin{equation}
    \mathcal{L}^H = \tilde{\Sigma}_{IJ}\wedge R^{IJ} - d\left(\tilde{\Sigma}_{IJ}\wedge \omega_s^{IJ}\right)
\end{equation}
Since the full curvature $R^{IJ}$ is a genuine $SO(1,3)$ tensor, its transformation rule is straightforward, and the product $\tilde{\Sigma}_{IJ}\wedge R^{IJ}$ is gauge-invariant. From (\ref{Eq:Trans2}) and (\ref{Eq:TransSigma}), we then have
\begin{equation}
    \mathcal{L}^{H\prime}=\mathcal{L}^H -d\left( \tilde{\Sigma}^\prime_{IJ}\wedge\alpha_s^{IJ}\right)
\end{equation}
Thus, the Herglotz Lagrangian changes by an exact form, and is gauge invariant at the level of the action, subject to the usual boundary assumptions. In general, Herglotz Lagrangians obey constrained variational principles, for the action 3-form $S$ is required to be such that $dS=\mathcal{L}^H$ \cite{herglotz1930lectures}. Note that, from (\ref{Eq:TransS}), the transformation rules we have deduced are consistent with the preservation of this constraint, as expected.\\

The behaviour of the components of the Herglotz theory under local Lorentz transformations is not quite sufficient to complete the physical interpretation of our theory. It is clear that the non-connection properties of $\omega_\perp$ and $\omega_s$ have significant implications for the way in which structures built from these objects behave under local $SO(1,3)$ transformations. However, it is not yet clear how these non-tensorial transformation laws are to be interpreted physically. For this, a necessary intermediate step is the calculation of the equations of motion.
\section{Dynamical Equations}\label{Sec:EoM}
We now compute the dynamics described by the Herglotz Lagrangian (\ref{Eq:HerglotzL2}). The action dependence of $\mathcal{L}^H$ means that we cannot simply integrate (\ref{Eq:HerglotzL2}) over the spacetime manifold, and take variations. Such a naive approach would lead to incorrect dynamics. Instead, the constraint $dS=\mathcal{L}^H$ must be enforced through a (scalar) Lagrange multiplier $\lambda$, and we should consider variations of the extended action
\begin{equation}\label{Eq:ExtendedAction}
    \widehat{S}=\int\bigl[\left(1-\lambda\right)dS+\lambda\mathcal{L}^H\bigl] 
\end{equation}
Variations with respect to the action 3-form $S$ determine the behaviour of the Lagrange multiplier. This is used when performing partial integration in subsequent calculations. We find that
\begin{equation}\label{Eq:lambda}
    d\lambda = \frac{\lambda}{3}\left(2\Theta+\star\,S\right)
\end{equation}
In order to vary (\ref{Eq:ExtendedAction}) with respect to the shape component $\omega_\perp$, we must pay careful attention to the fact that this too is a constrained calculation. Referring to the decomposition of the connection made in section (\ref{Sec:Decomp}), we recall that $\omega_\perp$ arose precisely as the kernel of the map $\Phi^{(1)}$. As such, $\tilde{\Sigma}_{IJ}\wedge\omega^{IJ}_\perp=0$, and this must be preserved with dynamical evolution. As such, we are obliged to augment the action (\ref{Eq:ExtendedAction}) with an additional Lagrange multiplier term of the form $\kappa\,\tilde{\Sigma}_{IJ}\wedge\omega^{IJ}_\perp$.\\

While, a priori, the insertion of this Lagrange multiplier is essential, when one carries out the variational calculation, it is found that $\kappa$ is forced to be identically zero. Thus, interestingly, the constraint $\tilde{\Sigma}_{IJ}\wedge\omega^{IJ}_\perp=0$ is enforced dynamically without need for our intervention. Introducing the quantity
\begin{equation}\label{Eq:Vartheta}
    \vartheta:=\frac{2}{3}\left(\Theta+\star\, S\right)
\end{equation}
the dynamical equation for the shape-projected component $\omega_\perp$ takes a beautifully simple form
\begin{equation}\label{Eq:EoM1}
    \boxed{D_\perp\tilde{\Sigma}_{IJ} +\vartheta\wedge\tilde{\Sigma}_{IJ}=0}
\end{equation}
As a means of verifying that our construction does in fact reproduce the true dynamics of the original system, it is productive to consider variations of the full Palatini action, expressed as in (\ref{Eq:FinalL}). The statement regarding the dynamical preservation of the condition $\tilde{\Sigma}_{IJ}\wedge\omega_\perp^{IJ}=0$ remains true here, and so we consider variations of the action
\begin{equation*}
    S_{\text{ext}} = \int e^\phi\,\Big[\tilde{\Sigma}_{IJ}\wedge R_\perp^{IJ}-d\phi\wedge\left(\tilde{\Sigma}_{IJ}\wedge\omega_s^{IJ}\right)-\left(D_\perp\tilde{\Sigma}_{IJ}\right)\wedge\omega_s^{IJ}+\tilde{\Sigma}_{IJ}\wedge\omega_s^I{}_K\wedge\omega_s^{KJ}+ \kappa\wedge\tilde{\Sigma}_{IJ}\wedge\omega_\perp^{IJ} \Big] 
\end{equation*}
Since this is the action for the full Palatini theory, and not the reduced Herglotz theory, we have independent variations to be carried out with respect to $\omega_s$ and $\omega_\perp$. We find that
\begin{align}
    \delta\omega_s\;&:\quad D_\perp\tilde{\Sigma}_{IJ} + d\phi\wedge\tilde{\Sigma}_{IJ} - 2\tilde{\Sigma}_{[I|K|}\wedge \omega^K_{s\,J]}=0\label{Eq:Var1}\\
    \delta\omega_\perp\;&:\quad D_\perp\tilde{\Sigma}_{IJ} + d\phi\wedge\tilde{\Sigma}_{IJ} - 2\tilde{\Sigma}_{[I|K|}\wedge \omega^K_{s\,J]}-\kappa\wedge\tilde{\Sigma}_{IJ} =0\label{Eq:Var2}
\end{align}
And so, as expected, the full Palatini theory also enforces $\tilde{\Sigma}_{IJ}\wedge\omega_\perp^{IJ}=0$ dynamically. However, this allows for a powerful consistency check to be carried out. Note that (\ref{Eq:Var2}) is a 3-form equation; this can be wedged with $\tilde{e}^I$, and the result written as the product of a scalar function and the spacetime volume form. With this, we deduce that the value of the (1-form) Lagrange multiplier $\kappa$ is
\begin{equation}
    \kappa=-\frac{2}{3}\Theta + d\phi -\frac{2}{3}\left(\iota_K\omega^K_{s\,J}\right)\tilde{e}^J =-\frac{2}{3}\Theta + d\phi -\frac{1}{3}\star S=  0
\end{equation}
where, in the penultimate equality, we have used the definition (\ref{Eq:Psi1}) of $\omega_s$ in terms of $\sigma=\star\,S$. Rearranging this expression for $d\phi$, we substitute this back into (\ref{Eq:Var1}) to yield
\begin{equation*}
    D_\perp\tilde{\Sigma}_{IJ} + \frac{1}{3}\left(2\Theta + \star\,S\right) \wedge\tilde{\Sigma}_{IJ} - 2\tilde{\Sigma}_{[I|K|}\wedge \omega^K_{s\,J]}=0
\end{equation*}
Finally, making use of the fact that
\begin{equation*}
    \tilde{\Sigma}_{[I|K|}\wedge \omega^K_{s\,J]} = -\frac{1}{6}\star S\wedge\tilde{\Sigma}_{IJ}
\end{equation*}
we find that the original Palatini equation of motion for $\omega_s$ may be expressed as
\begin{equation}
    D_\perp\tilde{\Sigma}_{IJ} +\frac{2}{3}\left(\Theta+\star\,S\right)\wedge\tilde{\Sigma}_{IJ}=0
\end{equation}
which is precisely (\ref{Eq:EoM1}), with $\vartheta=\frac{2}{3}\left(\Theta+\star\,S\right)$.
\section{The Role of Torsion}\label{Sec:Torsion}
Within the standard Palatini framework, variation of the action with respect to the full $SO(1,3)$ gauge connection gives rise to the following dynamical equation 
\begin{equation*}
    D\Sigma_{IJ}=0\qquad\implies\qquad T^I:= De^I = de^I+\omega^I_{\;K}\wedge e^K=0
\end{equation*}
Thus, on-shell, the torsion is vanishing. Returning to the Herglotz equation of motion for $\omega_\perp$
\begin{equation}\label{Eq:EoM2}
    D_\perp\tilde{\Sigma}_{IJ}+\vartheta\wedge\tilde{\Sigma}_{IJ}=0
\end{equation}
it is clear that we do not recover a dynamical torsion-free condition. In fact, the results of several preceding sections may now be used to provide the physical interpretation of our construction. In \cite{bell2026classical}, the way in which the scale encoded by the conformal mode was eliminated was somewhat different. Making the same decomposition $e^I=e^{\phi/2}\tilde{e}^I$ of the coframe, the connection was treated as follows. As described above, the dynamical equation $D\Sigma_{IJ}=0$, obtained from varying the Palatini action with respect to $\omega$, is equivalent to a torsion-free condition on the connection. It also, in some sense, defines $(e^I,\omega^{IJ})$ to be a `compatible pair', insofar as the covariant derivative (an object defined by $\omega^{IJ}$) annihilates the coframe $e^I$.\\

Upon making the conformal decomposition of the coframe, the torsion now reads
\begin{equation*}
    T^I=De^I=D\left(e^{\phi/2}\tilde{e}^I\right) = e^{\phi/2}\biggl[\frac{1}{2} d\phi\wedge\tilde{e}^I + d\tilde{e}^I + \omega^I_{\;K}\wedge\tilde{e}^K\biggl]
\end{equation*}
From this, it is clear that the dynamical equation $T^I=0$ certainly does \textit{not} define $(\tilde{e}^I,\omega^{IJ})$ to be compatible pair, since there is now an additional $d\phi\wedge\tilde{e}^I$ term. Thus, since $\tilde{e}^I$ is the coframe variable of our reduced theory, it is sensible to seek a connection $\tilde{\omega}$, such that variation of the Palatini action with respect to this new variable gives precisely the zero-torsion condition $\tilde{D}\tilde{e}^I:=d\tilde{e}^I+\tilde{\omega}^I_K\wedge\tilde{e}^K=0$. In this way, we will have a set $(\tilde{e}^I,\tilde{\omega}^{IJ})$ of scale-free compatible variables from which to construct the Herglotz theory. This was the physical motivation that, in \cite{bell2026classical}, led us to introduce
\begin{equation}\label{Eq:omegatilde}
    \tilde{\omega}^{IJ}:=\omega^{IJ}-\tilde{e}^{[I}\iota^{J]}d\phi
\end{equation}
Under a local Lorentz transformation, the additional piece $\tilde{e}^{[I}\iota^{J]}d\phi$ transforms tensorially, and so $\tilde{\omega}$ behaves as a genuine connection 1-form under internal $SO(1,3)$ transformations. The resulting Herglotz theory is, by construction, torsionless. This is precisely the conceptual step required to interpret the results of the present framework. The scale information was extracted in such a way that the remaining shape variable retained the full affine transformation law of a connection.\\

The torsion (or torsion-like properties) of the theory described by (\ref{Eq:HerglotzL2}) is governed by the 2-form $\Theta^I$. In the present work, we considered a decomposition $\omega=\omega_\perp+\omega_s$ of the model space $U_x$, which had the effect of projecting the affine term $\Lambda^{-1}d\Lambda$, as described by (\ref{Eq:Tranf}). Thus neither the shape nor the scale piece individually carries the full affine structure of the original connection. However, the crucial observation is that the component $\alpha_s$ of the affine translation that is missing, and thus obstructs $\omega_\perp$ from behaving as a connection, is precisely that which determines the inhomogeneous transformation properties of $\Theta^I$. This can easily be seen from (\ref{Eq:TransTheta}), where we deduced how $\Theta^I$ behaves under internal $SO(1,3)$ transformations. The conclusion is that the way in which the scale sector is removed determines whether the residual shape variable can retain the affine structure of a connection. If the scale contribution is removed tensorially, as in the construction in \cite{bell2026classical}, the residual variable remains a connection and can consistently be torsion-free. If instead the affine transformation law itself is decomposed between the projected sectors, the residual shape variable ceases to be a connection, and the missing affine structure manifests itself through an inhomogeneous transformation of the torsion-like sector.\\

There is a further component to our physical interpretation, which requires that we study in more detail the properties of $\Theta^I$. Let us make an algebraic decomposition of the shape projection $\omega_\perp^{IJ}$ into a reference value, denoted $\mathring{\omega}^{IJ}_\perp$, and an additional antisymmetric piece $K^{IJ}$. Thus, we write
\begin{equation}\label{Eq:omegaperpdecomp}
    \omega_\perp^{IJ}=\mathring{\omega}^{IJ}_\perp+K^{IJ}
\end{equation}
The defining property of $\mathring{\omega}_\perp$ is that $\mathring{D}\tilde{e}^I:=d\tilde{e}^I + \mathring{\omega}^I_{\perp K}\wedge \tilde{e}^K \equiv0$. Of course, our notation is intentionally highly suggestive that we ought to imagine the decomposition (\ref{Eq:omegaperpdecomp}) as a separation into a torsion-free piece $\mathring{\omega}_\perp^{IJ}$ and a contorsion tensor $K^{IJ}$. However, since the objects at our disposal do not have the correct transformation properties, this is only an analogy, and our decomposition is purely algebraic. Substitution into the dynamical equation (\ref{Eq:EoM2}) yields
\begin{equation}\label{Eq:ContorsionEoM}
    K^L_{\;I}\wedge\tilde{\Sigma}_{LJ} + K^L_{\;J}\wedge\tilde{\Sigma}_{IL} - \vartheta\wedge\tilde{\Sigma}_{IJ}=0
\end{equation}
While our decomposition must be considered exclusively algebraic in nature, $K^{IJ}$ is a perfectly well-defined $\Lambda^2V$ valued 1-form, and thus may be decomposed as the sum of a vector-trace, an axial-trace, and a trace-free piece. The subtlety is that $K^{IJ}$ will generally not transform tensorially. Thus, we write
\begin{equation}\label{Eq:KDecomp}
    K^{IJ} = V^{[I}\tilde{e}^{J]} + \frac{1}{2}\varepsilon^{IJ}_{\;\;\;KL}A^K\tilde{e}^L + q^{IJ}
\end{equation}
Substitution of this decomposition of $K^{IJ}$ into (\ref{Eq:ContorsionEoM}) reveals that 0-form coefficients $A^K$, associated with the axial-like trace, together with the trace-free 1-form $q^{IJ}$ are forced to be identically zero. The non-zero components are the $V^I$, which feature in the vector-like trace. We find that
\begin{equation}\label{Eq:ContorsionFinal}
    K^{IJ}=2\tilde{e}^{[I}\vartheta^{J]}\qquad\implies\qquad \omega^{IJ}_\perp=\mathring{\omega}_\perp^{IJ}+2\tilde{e}^{[I}\vartheta^{J]}
\end{equation}
Equations (\ref{Eq:KDecomp}) and (\ref{Eq:ContorsionFinal}) add a piece of physical insight which is central to the picture of Palatini gravity that we are assembling. From the derivation $D_\perp\tilde{e}^I$ of the coframe field, we extract the following relationship between $\Theta^I$ and $K^{IJ}$
\begin{equation}\label{Eq:Contorsion-Torsion}
    \Theta^I:=D_\perp\tilde{e}^I = d\tilde{e}^I + \omega^I_{\perp K}\wedge\tilde{e}^K \stackrel{(\ref{Eq:omegaperpdecomp})}{=} \underbrace{d\tilde{e}^I + \mathring{\omega}^I_{\perp K}\wedge\tilde{e}^K}_{\equiv\,0} + K^I_{\;K}\wedge\tilde{e}^K\quad\implies\quad \boxed{\Theta^I = K^I_{\;K}\wedge \tilde{e}^K}
\end{equation}
A priori, the contorsion-like object $K^{IJ}$ possesses up to 24 independent components, which are distributed amongst $4$ vector trace, 4 axial trace, and 16 traceless degrees of freedom. This is precisely the content of (\ref{Eq:KDecomp}). Physically, the axial and traceless components act to twist and deform the coframe geometry; this is made manifest through the torsion, related to $K^{IJ}$ via (\ref{Eq:Contorsion-Torsion}). The key insight is that these sectors have a non-angle-preserving effect on the underlying geometry, while the vector trace components act through pure scaling. The fact that the equation of motion (\ref{Eq:EoM2}) forces the axial trace and the traceless parts of $K^{IJ}$ to vanish means that the torsion is entirely packaged within the scale-changing vector trace degrees of freedom. We therefore conclude that our reduced theory exhibits a `shape rigidity', in which, despite having a non-zero torsion, the modes of this tensor that cause deformations of the conformal geometry are dynamically suppressed to zero. Substituting our expression (\ref{Eq:ContorsionFinal}) into the relation (\ref{Eq:Contorsion-Torsion}), we find that 
\begin{equation}
    \Theta^I+\frac{2}{3}\left(\Theta+\star \,S\right)\wedge\tilde{e}^I=0
\end{equation}
Taking the trace of this equation implies that $\Theta=-2\star S$, so that
\begin{equation}\label{Eq:ThetaITheta}
    \Theta^I+\frac{1}{3}\Theta\wedge\tilde{e}^I=0
\end{equation}
The physical interpretation of our construction is now substantially more complete. By extracting the scale sector through a projection of the full connection onto complementary shape and scale subspaces, the resulting shape component $\omega_\perp$ does not transform as a connection under the full local $SO(1,3)$ symmetry. Indeed, its transformation law contains only the shape projection $\alpha_\perp$ of the affine term $\Lambda^{-1}d\Lambda$, while the complementary contribution $\alpha_s:=(\Lambda^{-1}d\Lambda)_s$ is assigned to the scale sector. Consequently, $\omega_\perp$ fails to carry the full affine transformation law of the original connection. This failure has a corresponding manifestation in the torsion-like quantity $\Theta^I:=D_\perp\tilde{e}^I$. Under a local Lorentz transformation, $\Theta^I$ possesses the expected homogeneous tensorial contribution, but also acquires an inhomogeneous shift. Crucially, this shift is controlled precisely by the same scale-sector contribution $\alpha_s$ which is absent from the transformation law of $\omega_\perp$. Thus, the non-connection behaviour of the projected shape sector and the non-tensorial behaviour of $\Theta^I$ are not independent features of the construction: both arise from the redistribution of the affine structure of the full connection between the projected shape and scale sectors.\\

The dynamical equations further show that the resulting torsion-like behaviour is highly constrained. The Herglotz dynamics force $\Theta^I$ to be of pure vector-trace type, eliminating the remaining irreducible torsional components. Consequently, the surviving torsion-like degree of freedom cannot produce arbitrary distortions of the underlying conformal geometry: the axial and traceless modes, which would encode independent twisting and shape-deforming effects, are dynamically suppressed. the only surviving irreducible component is the vector-trace component, which is naturally associated with local changes of scale. In this sense, the reduced theory exhibits a form of `dynamical shape rigidity': the loss of the full affine structure in the projected shape sector is accompanied by a torsion-like response, whose dynamically allowed form is restricted precisely to the sector capable of encoding scale-like deformations.\\

Within the full Palatini theory, local rescalings of a frame provide a geometric mechanism by which changes of scale are represented as pure vector-trace torsion \cite{iosifidis2019scale}.\footnote{We would like to extend our thanks to Tomi Koivisto for drawing our attention to this. His remark has contributed to a substantial revision and improvement of the physical interpretation of our construction.} In this framework, the conformal decomposition explicitly isolates the local scale degree of freedom in $\phi$. In the reduced Herglotz theory, that degree of freedom is removed as an independent field, while the equations dynamically permit only a pure vector-trace torsion-like sector. This surviving sector therefore has precisely the geometrical character required to encode changes of local scale, without introducing independent deformations of the underlying conformal geometry.\\

The above interpretation is not merely an analogy; under a frame rescaling $e^I\rightarrow e^\lambda e^I$, for some scalar function $\lambda:M\rightarrow\mathbb{R}$, the torsion $T^I$ of the full theory undergoes a pure vectorial shift, proportional to $d\lambda\wedge e^I$. Consider now the zero-torsion condition for the conformally-decomposed coframe, expressed in the variables of the reduced theory
\begin{equation}\label{Eq:De^I}
    De^I=0\qquad\implies\qquad \frac{1}{2}d\phi\wedge\tilde{e}^I + \Theta^I + \frac{1}{6}\tilde{e}^I\wedge\star\,S=0
\end{equation}
Taking the trace of the right-hand expression yields the relationship $\star S=3\,d\phi -2\Theta$, and not only allows us to independently recover $\Theta=-2\star S$, as found above, but also implies that $\Theta=2\,d\phi$. Substituting this into the expression (\ref{Eq:ThetaITheta}) derived from the Herglotz equations of motion, we find that
\begin{equation}
    \Theta^I = -\frac{2}{3}d\phi\wedge\tilde{e}^I
\end{equation}
The torsion-like quantity $\Theta^I$ is not merely constrained to belong to the same irreducible sector as the torsion induced by a local frame rescaling (that of pure vector trace): on shell, it is explicitly determined by the derivative of the conformal factor itself. The reduced theory therefore retains the local scale information removed from the independent field content in a dynamically constrained vector-trace torsion-like sector.
\section{Integrable and Non-Integrable Sectors}\label{Sec:Integrable}
Here, we briefly discuss a feature of the Herglotz theory that is somewhat secondary to the narrative we wish to present, but is highly interesting nontheless. A more thorough development of these ideas is identified as future work. Temporarily considering the Herglotz Lagrangian (\ref{Eq:HerglotzL2}) as a standalone entity, independent of the Palatini action from which it was derived, we note that the decomposition of the connection using the maps $\Phi^{(1)}$ and $\Psi^{(1)}$ naturally identifies the scale sector with a geometric 1-form $\sigma$
\begin{equation}\label{Eq:Sigma}
    \omega_s^{IJ} = \frac{1}{3}\tilde{e}^{[I}\iota^{J]}\sigma
\end{equation}
However, this decomposition places no further geometrical restrictions on the 1-form $\sigma$. It is only when the physical context of the decomposition of the coframe field is introduced that the 1-form is forced to be exact. This is precisely the conclusion extracted from the series of equalities given above, where we showed that the zero-torsion condition $T^I=0$, when expressed in Herglotz variables, gave $\sigma=-d\phi$.\\

These observations imply that the Herglotz theory we have constructed is rather more general than that coming from purely gravitational dynamics. The Palatini theory is reproduced by the Herglotz formalism precisely when the 1-form $\sigma$ is exact. This indicates that the Herglotz theory possesses both integrable and non-integrable sectors, and that its solution space is therefore richer than that of Palatini gravity. It is not currently clear what, if any, physical significance may be attributed to the non-integrable sector, in which 1-form $\sigma$ is non-exact. It would appear that the additional structure arises as a result of the underlying multicontact geometry.
\section{Abstraction of the Decomposition}\label{Sec:Abstraction}
As we have discussed at length, the way in which the scale sector is excised to construct the Herglotz theory has implications for the nature of the objects of the reduced description. In particular, we observed that when the connection itself is projected onto disjoint shape and scale sectors, the resulting objects $\omega_\perp$ and $\omega_s$ fail to transform correctly as connection 1-forms. Further, this failure of the shape piece $\omega_\perp$ (which is the component of $\omega$ retained in the Herglotz theory) to behave as a connection may be traced to the fact that its transformation law lacks the full Maurer-Cartan term $\Lambda^{-1}d\Lambda$. The missing component $\alpha_s$ was found to control precisely the non-tensorial component of the transformation of the torsion-like form $\Theta^I$. The Herglotz equations of motion for our particular choice of decomposition maps gave
\begin{equation*}
    \Theta^I = -\frac{2}{3}d\phi\wedge\tilde{e}^I
\end{equation*}
While this \textit{is} of the form required to reproduce the effects of a frame rescaling, as described above, the coefficient $-2/3$ is not fixed by this physical/mathematical reasoning. Instead, this numerical factor may be attributed to the functional form of the Herglotz Lagrangian, which ultimately, may be traced to the precise algebraic form of the maps $\Phi^{(1)}$ and $\Psi^{(1)}$. We have previously advocated the physical viewpoint that the scale dynamics of the original theory must be recreated by a combination of the action dependence of $\mathcal{L}^H$ and the non-vanishing torsion, which we have now better understood in terms of frame rescalings. However, our observations now suggest that the precise way in which this burden is shared between $S$ and $\Theta^I$ is determined by the precise choice of decomposition maps.\\

In view of this, one might consider the problem of finding all possible independent decomposition maps. One may even be inclined to further conjecture the existence of a parameterisation of choice of decomposition of $\omega$. Intuitively, we imagine a dial, which allows us to tune how much of the scale dynamics is carried by the action density, and how much must be compensated by the internal torsion sector. In turning this dial, we pass through a family of Herglotz Lagrange densities, whose functional dependence on $S$ varies, but whose dynamical content always reproduces that of Palatini gravity (and possibly a more ample solution space).\\

More mathematically, we suppose that there exists a finite number $N$ of linear surjective morphisms $\Phi^{(a)}: U_x\longrightarrow  W_x$, with $\text{dim\,Ker}\,(\Phi^{(a)})=20$, whose right inverse $\Psi^{(a)}$ has an image of dimension 4. For every $a=1,\,\cdots,N$, the pair $(\Phi^{(a)}, \Psi^{(a)})$ is such that, once the functional form of the maps is specified, the following canonical decomposition is induced
\begin{equation}\label{Eq:Decomp}
    U_x= \text{Ker}(\Phi^{(a)}) \oplus\text{Im}(\Psi^{(a)})
\end{equation}
Our assumption of the existence of a finite $N$ is justified by the limited number of geometrical objects at our disposal, together with additional constraints on the set of allowed maps, which we now enumerate. Any surjection must contract all internal indices; of course, this does not require the resulting images $\Phi^{(a)}(\omega)$ to behave as scalars under internal $SO(1,3)$ transformations - we only require that all indices be contracted. Further, each map should be equivariant with respect to local Lorentz transformations, so that internal $SO(1,3)$ transformations of $\omega$ are appropriately reflected in the image $\Phi^{(a)}(\omega)$. Finally, the maps we wish to enumerate must be defined using only numerical constants, and the dynamical fields at our disposal: the coframe field $\tilde{e}^I$, the conformal mode $\phi$, and possibly its exterior derivative $d\phi$. These are then to be assembled using only the differential-geometric operations $(\wedge,\iota,d,\star)$.
\section{Finding Nemo (and his Lorentz-Equivariant Surjections)}\label{Sec:Nemo}
In order to enumerate the maps $\Phi^{(a)}$, it will be of use to employ representation-theoretic techniques. In order to justify the application of these methods to the space of gauge connections, we must distinguish the affine structure of the space of connections from the linear representation carried by their pointwise differences. Globally, the space $\mathcal{A}$ of $SO(1,3)$ connections is an affine space modeled on the vector space $\Omega^1(M,\frak{so}(1,3))$ of $\frak{so}(1,3)$ valued 1-forms. Correspondingly, at a fixed point $x\in M$, the fibre of the affine bundle of connections is an affine space modeled on the finite-dimensional vector space $U_x:=T_x^*M\otimes\Lambda^2V$. As discussed extensively, under a local Lorentz transformation $\Lambda$, the local representative of a connection transforms affinely according to
\begin{equation}
	\omega_x\longmapsto \text{Ad}_{\Lambda(x)^{-1}}\omega_x+\Lambda(x)^{-1} d\Lambda|_x.
\end{equation}
The inhomogeneous term is determined by the first jet $j_x^1\Lambda$ of the space $J^1(M, SO(1,3))$, and at the point $x$, is itself an ordinary element of the model vector space $U_x$. Hence the affine gauge transformation does not alter the underlying linear Lorentz representation carried by $U_x$; it merely acts on the affine space of connection values by a homogeneous Lorentz transformation together with a translation.\\

The $\Phi^{(a)}$ we wish to enumerate are linear maps on this model vector space, with codomain $W_x$. Lorentz equivariance is then understood with respect to the homogeneous linear representations of $SO(1,3)$ on $U_x$ and $W_x$, and so, schematically, we have
\begin{equation}\label{Eq:Equivariance}
    \Phi^{(a)}(\rho_U(\Lambda)\cdot\omega) = \rho_W(\Lambda)\cdot\Phi^{(a)}(\omega)
\end{equation}
in which $\rho_U$ and $\rho_W$ denote the appropriate $SO(1,3)$ representations for $U_x$ and $W_x$, respectively. The representation-theoretic classification therefore concerns $\text{Hom}_{SO(1,3)}(U_x,W_x)$, rather than the affine space of connections itself. In particular, the inhomogeneous term in the gauge transformation is not being discarded; rather, it is irrelevant to the question of enumerating Lorentz-equivariant linear maps on the model representation. If one instead demanded that $\Phi^{(a)}(\omega)$ itself be invariant under the full affine gauge transformation of a connection, an additional condition on the inhomogeneous term would have to be imposed. However, this is not something that we require; indeed, as we have seen in section (\ref{Sec:Trans}), the image $\Phi^{(1)}(\omega)$ is generally not invariant under local Lorentz transformations. Using the coframe to identify the relevant spacetime and internal Lorentz representations, the model space representation is (after complexification)
\begin{equation}\label{Eq:Irrep}
    \left(\frac{1}{2},\frac{1}{2}\right) \otimes \left[\left(1,0\right)\oplus\left(0,1\right)\right] = \left(\frac{3}{2},\frac{1}{2}\right) \oplus \left(\frac{1}{2},\frac{1}{2}\right)\oplus\left(\frac{1}{2},\frac{1}{2}\right)\oplus  \left(\frac{1}{2},\frac{3}{2}\right)
\end{equation}
while that of the co-domain is simply $(1/2,1/2)$. Since the domain space contains precisely two copies of the co-domain representation, Schur's lemma implies that $\text{dim\,Hom}_{SO(1,3)}(U_x,W_x)=2$. Thus, the conclusion is that there exist precisely two Lorentz-equivariant linear maps. Physically, the statement is that there are two, and only two, independent ways that one can choose to decompose the space of connections, in a way that extracts orthogonal shape and scale pieces, and reproduces the physical dynamics of Palatini gravity.
\section{A Moduli Space of Frictional Field Theories}\label{Sec:ModuliSpace}
Since the space of admissible equivariant maps is two-dimensional, we may select a basis $\{\Phi^{(1)},\Phi^{(2)}\}$, and consider the corresponding family of non-trivial linear combinations $\Phi_{(\xi,\eta)}:=\xi\Phi^{(1)}+\eta\Phi^{(2)}$, with $(\xi,\eta)\in\mathbb{R}^2\backslash\{0\}$. At each point $x\in M$, these maps act between the finite-dimensional representation spaces $U_x$ and $W_x$. Having established their pointwise form, we assume throughout that the corresponding bundle morphisms have been smoothly extended over M. Each $\Phi^{(a)}$ therefore induces a map on sections $\Phi^{(a)}:\Gamma(T^*M\otimes\Lambda^2V)\longrightarrow\Omega^3(M)$, and similarly for every linear combination $\Phi_{(\xi,\eta)}$. We do not bother to refine our notation to reflect this extension; however, we do denote the domain and codomain spaces of sections via $U$ and $W$, so as to distinguish them from their pointwise definitions in terms of finite-dimensional vector spaces. It should be understood that the representation-theoretic classification concerns the pointwise linear maps between the model spaces $U_x$ and $W_x$, rather than maps on the affine space of connections.\\

Linearity of the maps $\Phi^{(a)}$ alone is insufficient to guarantee that surjectivity is preserved for the non-trivial linear combinations $\Phi_{(\xi,\eta)}$ introduced above. There is, however, a representation-theoretic justification as to why all maps constructed are guaranteed to have the requisite properties described in section (\ref{Sec:Abstraction}). The key observation was fiven above: $\text{dim}\,\text{Hom}_{SO(1,3)}(U_x,W_x)=2$ is a statement that the model space representation $U_x$ contains exactly two copies of that of $W_x$. Thus, we write
\begin{equation}
    U_x\cong \left(W_x\otimes\mathbb{R}^2\right)\oplus K_x
\end{equation}
in which $\text{dim}\,K_x=16$, and $K_x$ contains all other irreducible components of $U_x$. Under this decomposition, an equivariant map $\Phi:U_x\longrightarrow W_x$ may be expressed on $W_x\otimes\mathbb{R}^2$ as $\text{id}_{W_x}\otimes \varphi$, for $\varphi\in\mathbb{R}^2\backslash\{0\}$, and vanishes on $K_x$. Let $\{e_1,e_2\}$ denote an abstract basis for the multiplicity space, with dual basis $\{e^1,e^2\}$, and write $\Phi^{(a)}=\text{id}_{W_x}\otimes e^a$, for each of the two independent maps. The arbitrary non-trivial linear combination $\Phi_{(\xi,\eta)}$ is represented as $\text{id}_{W_x}\otimes \left( \xi e^1 + \eta e^2\right)$. A non-zero linear functional $\xi e^1 + \eta e^2:\mathbb{R}^2\longrightarrow \mathbb{R}$, with a one-dimensional co-domain, is guaranteed to be surjective. This provides the representation-theoretic proof that any non-zero linear combination of surjective equivariant maps will also be surjective.\\

It is now possible to picture an abstract two-dimensional space of connection decomposition choices. Without loss of generality, we take $\Phi^{(1)}$ and $\Phi^{(2)}$ to be orthonormal with respect to a chosen inner product on the space of maps. Since $\Phi^{(1)}$ is constructed from the vector-like trace, the natural choice for the perpendicular direction is its axial counterpart. Thus, up to normalisation, we write
\begin{equation}\label{Eq:ChiralTrace}
    \Phi^{(2)}(\omega) := -\star\,\tilde{\Sigma}_{IJ}\wedge \omega^{IJ}
\end{equation}
We refer to the abstract two-dimensional space spanned by $(\Phi^{(1)},\Phi^{(2)})$ as the `parameter space of equivariant surjections', denoted $\mathbb{P}^{(\Phi)}$. Points of $\mathbb{P}^{(\Phi)}$ may be characterised by a pair of local coordinates $(\xi,\eta)$, corresponding to the map $\Phi_{(\xi,\eta)}=\xi\Phi^{(1)}+\eta\Phi^{(2)}$. At each point $x\in M$, this map has a kernel of dimension 20, and an inverse $\Psi_{(\xi,\eta)}$, whose image is of dimension 4. Once the local functional form of $\Phi_{(\xi,\eta)}$ has been specified, the decomposition of the model space $U_x$ is fixed and canonical with respect to that particular choice.\\

Note that $\mathbb{P}^{(\Phi)}$ is not simply $\text{Hom}_{SO(1.3)}(U,W)\cong\mathbb{R}^2$, since the space of maps which produce a valid connection decomposition categorically excludes $(\xi,\eta)=(0,0)$. This point gives rise to the zero map, whose kernel is 24-dimensional; in this case, the construction fails. Consequently, we have
\begin{equation}
    \mathbb{P}^{(\Phi)} = \bigl\{ \xi\Phi^{(1)}+\eta\Phi^{(2)}\,\in\,\text{Hom}_{SO(1,3)}(U,W)\;|\;(\xi,\eta)\in\mathbb{R}^2\backslash\{0\}\,\bigl\}\;\cong \mathbb{R}^2\backslash\{0\}\cong \mathbb{R}_+\times S^1
\end{equation}
When the connection $\omega=\omega_\perp+\omega_s$, determined by the decomposition $\text{Ker}(\Phi_{(\xi,\eta)}) \oplus\text{Im}(\Psi_{(\xi,\eta)})$, is introduced into the Palatini Lagrangian, application of the standard reduction procedure, in which the Lagrange density is expressed as $\mathcal{L} =e^\phi\mathcal{F}$, will lead to a Herglotz Lagrangian, for which both the functional dependence on the action 3-form $S$, and the torsional properties are determined by the coordinates $(\xi,\eta)$. Displacement within the parameter space corresponds to changing the algebraic structure of the Herglotz Lagrangian. Such structural changes are reflected in how the scale dynamics is distributed between the action-dependent and torsion-like sectors. The dynamics derived from a Herglotz Lagrangian at \textit{any} point in $\mathbb{P}^{(\Phi)}$ will always contain the dynamics of the Palatini theory as a subset. Further, if we restrict to the integrable sector, in which $\star\,S$ is exact, every point in $\mathbb{P}^{(\Phi)}$ gives rise to identical dynamics.\\

The parameter space contains significant redundancy, since the shape sector is determined by $\text{Ker}(\Phi_{(\xi,\eta)})$, which is invariant under $\mathbb{R}^\times:=\mathbb{R}\backslash\{0\}$ rescalings. The true moduli space, denoted $\mathbb{M}^{(\Phi)}$, is therefore
\begin{equation}\label{Eq:ModuliSpaceDef}
    \mathbb{M}^{(\Phi)}=\mathbb{P}^{(\Phi)}/\,\mathbb{R}^\times\cong \mathbb{RP}^1
\end{equation}
Elements of $\mathbb{M}^{(\Phi)}$ are equivalence classes $\Phi_{[\xi:\eta]}$, labelled by a projective pair $[\xi:\eta]\in \mathbb{RP}^1$. We refer to the corresponding homogeneous coordinates using round brackets $(\xi,\eta)$, and will always work with a particular homogeneous coordinate system determined by our basis maps. Explicitly, we write
\begin{equation}\label{Eq:Refinement}
    \Phi_{[\xi:\eta]}:=\bigg\{ \Phi_{(\lambda\xi,\lambda\eta)}=\lambda\left(\xi\Phi^{(1)}+\eta\Phi^{(2)}\right)\;|\;\lambda\in\mathbb{R}^\times\,\bigg\}
\end{equation}
With this, it follows that the moduli space $\mathbb{M}^{(\Phi)}$ may be written as
\begin{equation}
    \mathbb{M}^{(\Phi)}=\bigl\{\Phi_{[\xi:\eta]}\;\bigl|\; [\xi:\eta]\in\mathbb{RP}^1\,\bigl\}\;\cong \mathbb{RP}^1
\end{equation}
At each point $\Phi_{[\xi:\eta]}\in\mathbb{M}^{(\Phi)}$, there exists a distinguished one-dimensional vector subspace, namely the line generated by the projective point: $L_{[\xi:\eta]}=\text{span}\bigl\{\xi\Phi^{(1)}+\eta\Phi^{(2)}\bigl\}$. As $[\xi:\eta]$ varies over $\mathbb{M}^{(\Phi)}$, these one-dimensional subspaces fit together to form the tautological bundle $\mathcal{O}(-1)\longrightarrow\mathbb{RP}^1$. Abstractly, the total space $\mathcal{O}(-1)$ may be written as
\begin{equation}\label{Eq:O(-1)}
    \mathcal{O}(-1) = \bigl\{ \left([\xi:\eta],v\right)\in\mathbb{RP}^1\times\mathbb{R}^2\;\bigl|\; v\in\text{span}\{(\xi,\eta)\}\bigl\}
\end{equation}
There is then a canonical isomorphism $\iota: \mathbb{R}^2\longrightarrow \text{Hom}_{SO(1,3)}(U,W)$, which acts in the obvious manner, taking $(\xi,\eta)\longmapsto\xi\Phi^{(1)}+\eta\Phi^{(2)}$. Under this isomorphism, the total space becomes 
\begin{equation}\label{Eq:O(-1)2}
    \iota\left(\mathcal{O}(-1)\right) = \bigg\{ \left(\Phi_{[\xi:\eta]}, \lambda\left(\xi\Phi^{(1)}+\eta\Phi^{(2)}\right)\right) \in\mathbb{M}^{(\Phi)}\times \text{Hom}_{SO(1,3)}(U,W)\;\bigl|\;  [\xi:\eta]\in\mathbb{RP}^1\;,\;\lambda\in\mathbb{R}\bigg\}
\end{equation}
Throughout, it is understood that referring to a `point' $[\xi:\eta]$ in $\mathbb{M}^{(\Phi)}$ is synonymous with the map $\Phi_{[\xi:\eta]}$. Where potential cause for confusion may arise, we will be more careful. The key point is that the isomorphism $\iota$ establishes a geometrical correspondence between the abstract space $\mathcal{O}(-1)\subset \mathbb{RP}^1\times\mathbb{R}^2$, and the concrete space of surjective intertwining maps; in particular $\iota\left(\mathcal{O}(-1)\right)\subset \mathbb{M}^{(\Phi)}\times\text{Hom}_{SO(1,3)}(U,W)$. From (\ref{Eq:O(-1)2}), the full tautological bundle clearly contains the zero-section $s_0:\mathbb{RP}^1\longrightarrow\mathcal{O}(-1)$, sending $\Phi_{[\xi:\eta]}\longmapsto \left(\Phi_{[\xi:\eta]},0\right)$. The zero map is not included in the admissible parameter space $\mathbb{P}^{(\Phi)}$, and so the physically relevant bundle is $\mathcal{O}(-1)^\times:=\mathcal{O}(-1)\backslash s_0$.\\

Turning to the evaluation of the linear maps defined by points of $\mathbb{M}^{(\Phi)}$, recall that, by construction, $\Phi^{(1)}$ and $\Phi^{(2)}$ form an orthonormal basis of $\text{Hom}_{SO(1,3)}(U,W)$ with respect to a chosen inner product. Despite this, linear independence of the corresponding images $\Phi^{(1)}(\omega)$ and $\Phi^{(2)}(\omega)$ does not immediately follow. Let $\mathcal{A}$ denote the infinite-dimensional affine space of $SO(1,3)$ gauge connections. Since the flat connection $\omega=0$ maps to the zero 3-form, which guarantees the failure of $\Phi^{(1)}(\omega)$ and $\Phi^{(2)}(\omega)$ to be linearly independent, we restrict the domain of the linear maps to the punctured space $\mathcal{A}_0:=\mathcal{A}\backslash\{\omega=0\}$. Note, however, that exclusion of $\omega=0$ from the domain is not, a priori, a sufficient condition to prevent the images becoming linearly dependent. In order to evaluate a map at a given $\omega\in\mathcal{A}_0$, we introduce the linear map $ev_\omega:\text{Hom}_{SO(1,3)}(U,W)\longrightarrow \Omega^3(M)$, with $ev_\omega(\Phi):=\Phi(\omega)$. Since $\iota(\mathcal{O}(-1))\subset \mathbb{M}^{(\Phi)}\times \text{Hom}_{SO(1,3)}(U,W)$, we can restrict $ev_\omega$ to the tautological fibres, and introduce the bundle morphism
\begin{equation}\label{Eq:Eval}
    \begin{split}
        E_\omega \;:\; \iota(\mathcal{O}(-1))\,&\longrightarrow\, \mathbb{M}^{(\Phi)}\times \Omega^3(M)\\
        \left(\Phi_{[\xi:\eta]}, \lambda\left(\xi\Phi^{(1)}+\eta\Phi^{(2)}\right)\right) \,&\longmapsto\, \left(\Phi_{[\xi:\eta]},\lambda\left(\xi\Phi^{(1)}(\omega)+\eta\Phi^{(2)}(\omega)\right)\right)
    \end{split}
\end{equation}
Since the space of maps of interest does not include the zero map, we restrict $E^\times_\omega:=E_\omega\left.\right|_{\iota(\mathcal{O}(-1)^\times)}$, which guarantees that $\lambda\in\mathbb{R}^\times$. Suppose now that we choose the images to be of the form
\begin{equation}\label{Eq:2maps}
    \Phi^{(1)}(\omega)=-\tilde{\Sigma}_{IJ}\wedge \omega^{IJ}\qquad\qquad \Phi^{(2)}(\omega)=-\star\tilde{\Sigma}_{IJ}\wedge\omega^{IJ}
\end{equation}
then everywhere-non-degeneracy of the coframe guarantees that, for a given connection $\omega\in\mathcal{A}_0$, the space $V_\omega:=\text{span}\{\Phi^{(1)}(\omega),\Phi^{(2)}(\omega)\}\subset \Omega^3(M)$ is two-dimensional. In particular, this means that the map $E^\times_\omega$ does not collapse the $\mathcal{O}(-1)$ fibre. If we define an `image bundle' 
\begin{equation}
    \mathcal{E}^\times_\omega:= \bigg\{\left(\Phi_{[\xi:\eta]} , \lambda\left(\xi\Phi^{(1)}(\omega)+\eta\Phi^{(2)}(\omega)\right)\right)\;\biggl|\; [\xi:\eta]\in\mathbb{RP}^1\;,\; \lambda\in\mathbb{R}^\times\bigg\}
\end{equation}
then $E^\times_\omega:\iota(\mathcal{O}(-1)^\times)\longrightarrow \mathcal{E}^\times_\omega$ is an isomorphism of bundles over $\mathbb{M}^{(\Phi)}$. Note that there may exist choices of basis maps for which linear independence of the images is not guaranteed for all $\omega\in\mathcal{A}_0$. This is not the case for the physically-motivated choice (\ref{Eq:2maps}); however, in order to make our construction independent of this assumption, we should further restrict $\mathcal{A}_0$ to the open set
\begin{equation}
    \widehat{\mathcal{A}}:=\bigl\{\omega\in\mathcal{A}_0\;\bigl|\;\xi\Phi^{(1)}(\omega)+\eta\Phi^{(2)}(\omega)=0\;\;\implies\;\;\xi=\eta=0\bigl\}
\end{equation}
With this, the construction goes through as above, for a generic choice of maps.
\section{A Bundle Over $\mathcal{A}_0$}\label{Sec:Bundle}
In order to understand the effects of allowing $\omega$ to vary over $\mathcal{A}_0$, we introduce\footnote{In the case that the basis maps chosen are not those of (\ref{Eq:2maps}), it is understood that all instance of $\mathcal{A}_0$ are to be replaced with $\widehat{\mathcal{A}}$.}
\begin{equation}\label{Eq:TotalSpace}
    \mathcal{P}^\times:= \bigl\{ \left(\omega,[\xi:\eta],\beta\right)\;\bigl|\;\beta\in L_{[\xi:\eta]}(\omega)^\times\bigl\}
\end{equation}
in which $L_{[\xi:\eta]}(\omega):=\text{span}\{\xi\Phi^{(1)}(\omega)+\eta\Phi^{(2)}(\omega)\}$ is the line of 3-forms defined by $[\xi:\eta]$, and $L_{[\xi:\eta]}(\omega)^\times:=L_{[\xi:\eta]}(\omega)\backslash\{0\}\cong\mathbb{R}^\times$ discards the zero 3-form. We propose that $\mathcal{P}^\times\longrightarrow \mathcal{A}_0\times\mathbb{M}^{(\Phi)}$ is an $\mathbb{R}^\times$ principal bundle, with fibre $L_{[\xi:\eta]}(\omega)^\times\cong\mathbb{R}^\times$, and projection $\pi:\mathcal{P}^\times\longrightarrow\mathcal{A}_0\times\mathbb{M}^{(\Phi)}$, that acts as $\pi(\omega,[\xi:\eta],\beta)= (\omega,[\xi:\eta])$. The free and transitive $\mathbb{R}^\times$ action on the fibres, for $q\in\mathbb{R}^\times$, is given by 
\begin{equation}
    \left(\omega,[\xi:\eta],\beta\right)\cdot q = \left(\omega,[\xi:\eta],q\beta\right)
\end{equation}
Motivated by this construction, consider, for each $\omega\in\mathcal{A}_0$, the two-dimensional vector space $V_\omega:=\text{span}\{\Phi^{(1)}(\omega),\Phi^{(2)}(\omega)\}\subset\Omega^3(M)$. Because of the linear independence of $\Phi^{(1)}(\omega)$ and $\Phi^{(2)}(\omega)$, we have a smooth rank-2 vector bundle 
\begin{equation}
    V:=\bigsqcup_{\omega\in\mathcal{A}_0} V_\omega\;\longrightarrow \; \mathcal{A}_0
\end{equation}
More concretely, this is a rank-2 vector subbundle of $\mathcal{A}_0\times\Omega^3(M)$, which since we have a smooth global frame $\{\Phi^{(1)}(\omega),\Phi^{(2)}(\omega)\}$, is trivial. Further, there exists an isomorphism $V\cong \mathcal{A}_0\times\mathbb{R}^2$, where a point $(\omega,\xi,\eta)\in\mathcal{A}_0\times\mathbb{R}^2$ is mapped to $\left(\omega,\xi\Phi^{(1)}(\omega)+\eta\Phi^{(2)}(\omega)\right)\in V$. It follows that the projectivisation $\mathbb{P}(V)$ satisfies $\mathbb{P}(V)\cong\mathcal{A}_0\times\mathbb{RP}^1$, with elements of $\mathcal{A}_0\times\mathbb{RP}^1$ being of the form $(\omega,[\xi:\eta])$, which we trivially map to $(\omega,\Phi_{[\xi:\eta]})$. We thus have 
\begin{equation*}
    \mathbb{P}(V)\cong \mathcal{A}_0\times\mathbb{RP}^1\cong \mathcal{A}_0\times\mathbb{M}^{(\Phi)}
\end{equation*}
For the tautological bundle over $\mathbb{P}(V)$, denoted $\mathcal{O}_{\mathbb{P}(V)}(-1)\longrightarrow \mathbb{P}(V)$, elements of the total space are of the form $(\omega,[\xi:\eta],v)$, with $v\in L_{[\xi:\eta]}(\omega)$. Removing the zero-section, it should be clear that $\mathcal{O}_{\mathbb{P}(V)}(-1)^\times\longrightarrow \mathbb{P}(V)\cong\mathcal{A}_0\times\mathbb{M}^{(\Phi)}$ is isomorphic to $\mathcal{P}^\times\longrightarrow \mathcal{A}_0\times\mathbb{M}^{(\Phi)}$. In fact, we have a somewhat stronger statement: $\mathcal{P}^\times$ is the concrete realisation of the punctured tautological bundle $\mathcal{O}_{\mathbb{P}(V)}(-1)^\times$, once we have made the identification $\mathbb{P}(V)\cong \mathcal{A}_0\times\mathbb{M}^{(\Phi)}$.\\

Having established that $\mathcal{P}^\times\cong \mathcal{O}_{\mathbb{P}(V)}(-1)^\times$, it is straightforward to see that $\mathcal{P}^\times$ is non-trivial. Introduce the map $\text{pr}_2:\mathcal{A}_0\times\mathbb{RP}^1\longrightarrow\mathbb{RP}^1$, acting by projecting to the second factor. We then have 
\begin{equation*}    
    \mathcal{O}_{\mathbb{P}(V)}(-1)\cong\text{pr}_2^*\mathcal{O}_{\mathbb{RP}^1}(-1)\qquad\implies\qquad \mathcal{P}^\times\cong\text{pr}_2^*\left(\mathcal{O}_{\mathbb{RP}^1}(-1)^\times\right)
\end{equation*}
Thus $\mathcal{P}^\times$ is (isomorphic to) the pullback bundle, and this is non-trivial precisely because its restriction to a particular slice $\{\omega_0\}\times\mathbb{RP}^1$ is just $\mathcal{O}_{\mathbb{RP}^1}(-1)^\times$, which is non-trivial.\\

Summarising the results of our study of the moduli space, with a focus on their physical significance, we found that $\mathbb{M}^{(\Phi)}\cong\mathbb{RP}^1$, so that its points are characterised by a pair of projective coordinates $[\xi:\eta]$, corresponding to an equivalence class of linear maps. As $[\xi:\eta]$ varies over $\mathbb{M}^{(\Phi)}$, the linear subspaces $L_{[\xi:\eta]}=\text{span}\bigl\{\xi\Phi^{(1)}+\eta\Phi^{(2)}\bigl\}$ fit together to form the tautological bundle $\iota(\mathcal{O}(-1))\longrightarrow \mathbb{M}^{(\Phi)}$. Movement along the fibres changes the representative of the point, but does not affect the decomposition or resulting Herglotz Lagrangian. Movement in the base, however, takes us to a new equivalence class of maps, inducing an alternative connection decomposition. Physically, this reorganises the way in which the scale dynamics are distributed between the action 3-form and the torsion-like sector.\\

The maps in $\mathbb{M}^{(\Phi)}$ act on connection 1-forms to produce spacetime 3-forms. For a fixed $\omega\in\mathcal{A}_0$, there exists a natural evaluation map $ev_\omega:\text{Hom}_{SO(1,3)}(U,W)\longrightarrow \Omega^3(M)$, acting as $\Phi\longmapsto\Phi(\omega)$. When restricted fibrewise to the bundle $\iota(\mathcal{O}(-1))$, evaluation at different points along the fibre produces a line $L_{[\xi:\eta]}(\omega)^\times$ of 3-forms within $\Omega^3(M)$. All action 3-forms along this line are equivalent at the level of the Herglotz Lagrangian. From this, we constructed an $\mathbb{R}^\times$ bundle $\mathcal{P}^\times\longrightarrow \mathcal{A}_0\times \mathbb{M}^{(\Phi)}$, with fibre $L_{[\xi:\eta]}(\omega)^\times$. Movement in the fibre direction changes only our position along the line of equivalent 3-forms. Displacement in the base, however, changes both the equivalence class of maps \textit{and} at which point in $\mathcal{A}_0$ they are evaluated.\\

In order to separate changes due to movement in the connection space from those caused by displacements within $\mathbb{M}^{(\Phi)}$, we introduced the trivial vector bundle $V\longrightarrow\mathcal{A}_0$. Movement in $\mathcal{A}_0$ alters the two-dimensional subspace $\text{span}\{\Phi^{(1)}(\omega),\Phi^{(2)}(\omega)\}\subset\Omega^3(M)$, but is independent of position in the moduli space. Physically, this bundle measures how our chosen basis maps respond to a change of connection 1-form. This does not alter the functional form of the Herglotz Lagrangian, as this depends only on the maps chosen, and our position in the moduli space. However, changing $\omega$ does alter the physical dynamics, since the gauge connection is a dynamical field that enters the variational calculation of the equations of motion.
\section{Summary and Discussion}
In this work, we have systematically extended the contact reduction of first-order Palatini gravity by moving beyond the physically intuitive picture of \cite{bell2026classical} into a rigorous representation-theoretic framework. By mapping the problem of the decomposition of the $SO(1,3)$ gauge connection to that of enumerating the Lorentz-equivariant intertwining maps, we were able to use Schur’s lemma to establish that there exist precisely two independent decomposition choices. This observation naturally led to a smooth, one-dimensional moduli space of frictional field theories $\mathbb{M}^{(\Phi )}$. Topologically, this space was identified with the real projective line $\mathbb{RP}^1$, and so could be parameterised with a pair of projective coordinates $[\xi:\eta]$.\\

In the first part of our work, we explicitly constructed the Herglotz Lagrangian for a particular choice of connection decomposition. In the language of the moduli space $\mathbb{M}^{(\Phi)}$, our choice was characterised by the projective coordinates $[1:0]$. Since the extraction of the scale sector of the full theory was carried out at the level of the connection, we found that the individual projected components $\omega_\perp$ and $\omega_s$ did not behave as connection 1-forms. However, the Herglotz equations of motion revealed that this particular Lagrangian gave rise to non-vanishing torsion. Indeed, by examining the dynamical equations in detail, we argued that the presence of a torsion-like sector should be a general feature of any Herglotz Lagrangian obtained from a decomposition of this type. This claim was motivated principally by the observation that the torsion was highly constrained: all torsional modes capable of deforming the underlying conformal geometry were dynamically forced to zero. The form of $\Theta^I$, when expressed in the variables of the original palatini theory, was found to be of precisely the correct form to mimic a frame rescaling (in the sense of \cite{iosifidis2019scale}).\\

The physical interpretation of the moduli space discussed above was that each point gives an equivalence class of morphisms. At each spacetime point, these morphisms act as linear maps between two finite-dimensional vector spaces, providing a pointwise decomposition of the model vector space of connection values. Different points in the moduli space lead to Herglotz Lagrangians which all reproduce the dynamics of first-order Palatini gravity, but the way in which the scale-related physics is recreated is a decomposition-dependent combination of action dependence and non-vanishing torsion.\\

The framework presented opens up several interesting avenues for future study. For instance, our discovery of a non-integrable sector within the standalone Herglotz theory suggests that contact geometry naturally accommodates variations of gravity broader than standard Palatini dynamics. Our analysis was somewhat brief, and so further exploring whether these non-integrable configurations could be realised within a physical theory remains an open and interesting question. Further, we noted and discussed in detail that the Herglotz theory described by (\ref{Eq:HerglotzL2}) exhibited torsion-like effects. It would be of interest to conduct a more detailed mathematical study of these torsional configurations, examining, for example, their implications for spacetime geometry and geodesic motion.
\appendix
\section{Notation and Conventions}\label{App:A}
In this appendix, we compile a short list of the conventions used throughout. These are principally conventions related to differential forms in four dimensions. In general, a differential $p$-form in spacetime $\xi\in\Omega^p(M)$ will be normalised as
\begin{equation}\label{Eq:pform}
    \xi=\frac{1}{p!}\xi_{\mu_1\cdots\mu_p}\,dx^{\mu_1}\wedge\cdots\wedge dx^{\mu_p}
\end{equation}
The natural isomorphism $\Omega^p(M)\cong\Omega^{4-p}(M)$ is provided by the Hodge star operator $\star$, and for the $p$-form $\xi$, we have
\begin{equation}
    \star\,\xi = \frac{|\text{det}(e)|}{(4-p)!}\varepsilon_{\qquad\mu_1\cdots\mu_{4-p}}^{\nu_1\cdots\nu_p}\xi_{\nu_1\cdots\nu_p}\,dx^{\mu_1}\wedge\cdots \wedge dx^{\mu_{4-p}}
\end{equation}
Since our metric is Lorentzian, and we work in dimension four, we have $\star^2 = \text{id}$. Additionally, for two forms $\alpha\in\Omega^p(M)$ and $\beta\in\Omega^q(M)$, we will make frequent use of the following graded generalisation of the Leibniz rule
\begin{equation}
    \iota_X\left(\alpha\wedge\beta\right) = \iota_X\alpha\wedge\,\beta+ (-1)^p\,\alpha\wedge\,\iota_X\beta,\qquad\text{for any}\;X\in\mathfrak{X}^{\infty}(M)
\end{equation}
The totally antisymmetric Levi-Civita symbol $\varepsilon_{\mu\nu\kappa\rho}$ is such that, in all coordinate systems, we have $\varepsilon_{0123}=\varepsilon^{0123}=+1$. Indices of $\varepsilon_{\mu\nu\kappa\rho}$ are raised and lowered using the spacetime metric. On the flat internal space $(V,\eta)$, we have $\varepsilon_{IJKL}$, and we take $\varepsilon^{0123}=-\varepsilon_{0123}=1$. We also have the following useful identities for the contraction of two Levi-Civita symbols
\begin{align}
    \varepsilon^{IJKL}\varepsilon_{IMNP} &= -6\delta^{[J}_M\delta^K_N\delta^{L]}_P\\
    \varepsilon^{IJKL}\varepsilon_{IJMN} &=-4\delta^{[K}_M\delta^{L]}_N \\
    \varepsilon^{IJKL}\varepsilon_{IJKM} &=-6\delta^L_M
\end{align}
In the above expressions and throughout, we adopt the following conventions with respect to index (anti)symmetrisation
\begin{align}
    A^{[IJ]} &=\frac{1}{2}\left(A^{IJ}-A^{JI}\right)\\
    A^{[IJK]} &= \frac{1}{3}\left(A^{[IJ]K}+A^{[JK]I}+A^{[KI]J}\right)\\
    A^{[I|J|K]} &=\frac{1}{2}\left(A^{IJK}-A^{KJI}\right)
\end{align}
with similar expression for symmetrisation, replacing $[\;\;]$ with $\left(\,\,\right)$.\\

The spacetime volume form is denoted $\text{vol}$, and we have
\begin{equation}\label{Eq:Vol}
    \text{vol} = |\text{det}(e)|\,dx^0\wedge\, dx^1\wedge\, dx^2\wedge\, dx^3 = \frac{|\text{det}(e)|}{4!}\,\varepsilon_{IJKL}\,e^I\wedge e^J\wedge e^K\wedge e^L
\end{equation}
Additionally, the coframe determinant $\text{det}(e)$ satisfies the following relation
\begin{equation}
    \varepsilon_{\mu\nu\kappa\lambda} = |\text{det}(e)|\,e_\mu^{\;I}e_\nu^{\;J}e_\kappa^{\;K}e_\lambda^{\;L}\,\varepsilon_{IJKL}
\end{equation}
\section{Herglotz Lagrangians via Differential Forms}\label{App:B}
In previous work, we established the field-theoretic contact reduction process for multisymplectic Lagrangian functions of the form \cite{DSMulti}
\begin{equation}\label{Eq:e^phiLag}
    L(\phi,\partial_\mu\phi,\psi,\partial_\mu\psi)=e^\phi f(\partial_\mu\phi,\psi,\partial_\mu\psi)
\end{equation}
where $\phi$ is the scaling variable, and the dependence of $f$ on this field is exclusively through its coordinate derivatives $\partial_\mu\phi$. We use $\psi$ to denote an unscaled field, where for multiple such fields, may be indexed these as $\psi^a$, for example. For these Lagrangians, all calculations were performed very explicitly in a local chart of coordinates $x^\mu$, with all spacetime indices displayed. Since the contents of the first-order Palatini theory are best expressed in terms of differential forms, it will be of benefit to dedicate some effort to re-expressing some of the ideas developed in \cite{DSMulti} in a language more adapted to our objective. We know, for example, that the components $\sigma^\mu$ of the 1-form dual to $S$, and the Herglotz Lagrangian function $L^H$ may be found from (\ref{Eq:e^phiLag}) via
\begin{equation}\label{Eq:Old}
    \sigma^\mu :=\frac{\partial f}{\partial(\partial_\mu\phi)}, \qquad L^H:=f-(\partial_\mu\phi)\sigma^\mu
\end{equation}
The obvious generalisation is to consider a Lagrangian density $\mathcal{L}$ on the $d$-dimensional manifold $M$, constructed from differential forms
\begin{equation}\label{Eq:e^phiDensity}
    \mathcal{L}= e^\phi\mathcal{F}(d\phi,\psi,d\psi)
\end{equation}
$\mathcal{F}$ is a top-form on $M$, $\phi$ continues to denote a scalar field, and $\psi$ may now be a form of arbitrary degree $p\leq d$. The Herglotz constraint, which in \cite{DSMulti}, was given as $\partial_\mu\sigma^\mu=L^H$, is immediately generalised to $dS=\mathcal{L}^H$, with $\sigma=\star\, S$.\\

From the above construction, it is clear that we want to define our action density $S$ through some expression that, schematically, we write as $\partial \mathcal{F}/\partial(d\phi)$. The goal is to give rigorous mathematical meaning to this statement. Throughout, $M$ is assumed oriented (so that the integration of $d$-forms is well-defined), and all field variations are taken compactly supported in the interior of $M$, so that boundary terms may be neglected.\\

At a particular point $x\in M$, since $d\phi(x)\in\Lambda^1(T^*_xM)$, and similarly for the other arguments, we know that
\begin{equation*}
    \mathcal{F}\;:\;\Lambda^1(T_x^*M)\times\Lambda^p(T^*_xM)\times\Lambda^{p+1}(T_x^*M)\longrightarrow\Lambda^d(T_x^*M).
\end{equation*}
Note that the map $\mathcal{F}$ is built canonically from its arguments, and so is functionally identical at all points. In order to define differentiation with respect to a form, we prove an intermediate lemma. All differentiation is performed with $x$ held fixed, so that we may work with ordinary derivatives of maps between finite-dimensional real vector spaces.
\begin{lemma}
Let $V$ be a $d$-dimensional real vector space and $V^*$ its dual. For every linear map
\begin{equation*}
    T:\Lambda^p(V^*)\longrightarrow\Lambda^d(V^*)
\end{equation*}
there exists a unique $\beta_T\in\Lambda^{d-p}(V^*)$ such that
\begin{equation}
    T(\alpha)=\alpha\wedge\beta_T
    \qquad\text{for all }\alpha\in \Lambda^p(V^*)
\end{equation}
\end{lemma}
We leave the proof of this lemma, which is a standard exercise in linear algebra, to the reader. Considering the space $V=T_xM$, we apply the lemma to the linear map
\begin{equation*}
    \delta(d\phi)\longmapsto \left.\frac{d}{d\epsilon}\right|_{\epsilon=0}\mathcal{F}(d\phi+\epsilon\,\delta(d\phi),\psi,d\psi)
\end{equation*}
obtained from $\mathcal{F}$ by differentiating the $d\phi$ slot only, holding $\psi$ and $d\psi$ fixed. This is the ordinary directional derivative of a map between the finite-dimensional vector spaces $\Lambda^{1}(T_x^*M)$ and $\Lambda^d(T_x^*M)$, and so by our lemma, is represented by wedging against a unique $(d-1)$-form, which we define to be $S=S(d\phi,\psi,d\psi)$
\begin{equation}\label{Eq:DefS}
    \left.\delta \mathcal{F}\right|_{\psi,d\psi \,{\text{fixed}}} =\delta(d\phi)\wedge S \qquad \text{for every }\,\delta(d\phi)\in\Lambda^1(T_x^*M)
\end{equation}
Similarly, we take $Q=Q(d\phi,\psi,d\psi)$ to be the unique $(d-p)$-form with
\begin{equation}
    \left.\delta \mathcal{F}\right|_{d\phi,d\psi\, \text{fixed}} =  \delta \psi\wedge Q
\end{equation}
Finally, for the unique $(d-p-1)$-form obtained from application of the lemma to differentiation in the $d\psi$ slot, we introduce $P=P(d\phi,\psi,d\psi)$ 
\begin{equation}
    \left.\delta \mathcal{F}\right|_{d\phi,\psi\,\text{fixed}} =\delta(d\psi)\wedge P
\end{equation}
This construction gives us a well-defined notion of differentiation with respect to a form, and we henceforth employ the notation 
\begin{equation}\label{Eq:DefS}
    S=\frac{\partial\mathcal{F}}{\partial(d\phi)}
\end{equation}    
to signify that $S$ is the unique $(d-1)$-form, which satisfies (\ref{Eq:DefS}).\\

In order to complete the re-expression of the reduction procedure, we must assume that the second fibre derivative of $\mathcal{F}$ in the $d\phi$-direction is non-degenerate. By the implicit function theorem, $d\phi$ may then locally be written as a function $d\phi(\psi,d\psi,S)$. Referring to (\ref{Eq:Old}), the Herglotz Lagrangian is exactly what one would expect
\begin{equation}
    \mathcal{L}^H(\psi,d\psi,S):=\mathcal{F}\bigl(d\phi(\psi,d\psi,S),\psi,d\psi\bigr) -d\phi(\psi,d\psi,S)\wedge S
\end{equation}
With this, we are able to express the Herglotz-Lagrange equations exclusively in terms of differential forms:
\begin{equation}
    d\left(\frac{\partial \mathcal{L}^H}{\partial(d\psi)}\right) - (-1)^p \frac{\partial \mathcal{L}^H}{\partial \psi}=(-1)^d\left(\frac{\partial \mathcal{L}^H}{\partial(d\psi)}\right)\wedge\left(\frac{\partial \mathcal{L}^H}{\partial S}\right)
\end{equation}
We therefore have a construction which generalises the ideas of \cite{DSMulti}. From that work, we know that the Herglotz Lagrange density $\mathcal{L}^H$ reproduces the dynamics of the original system, but does so in a manner that makes no reference to the scaling variable $\phi$.
\bibliography{Refs}
\bibliographystyle{unsrt}
\end{document}